\documentclass[journal]{IEEEtran}
\usepackage{color}
\usepackage{subfigure}
\usepackage{amssymb}
\usepackage{amsmath}
\usepackage{algorithmic}
\usepackage{graphicx}
\usepackage{array}
\usepackage{url}
\usepackage{cite}
\usepackage{booktabs}

\DeclareMathOperator*{\maximize}{maximize}
\DeclareMathOperator{\sbto}{subject~to}

\begin{document}

\title{Deep Learning based Wireless Resource Allocation with Application to Vehicular Networks}
\author{Le~Liang,~\IEEEmembership{Member,~IEEE,}
        Hao~Ye,~\IEEEmembership{Student~Member,~IEEE,}\\
        Guanding~Yu,~\IEEEmembership{Member,~IEEE,}
        and~Geoffrey~Ye~Li~\IEEEmembership{Fellow,~IEEE}
\thanks{L. Liang is with Intel Labs, Hillsboro, OR, 97124 USA (e-mail: lliang@gatech.edu).}
\thanks{
H. Ye and G. Y. Li are with the School of Electrical and Computer Engineering, Georgia Institute of Technology, Atlanta,
GA, 30332 USA (e-mail: yehao@gatech.edu;liye@ece.gatech.edu).} 
\thanks{G. Yu is with Department of Information Science and Electronic Engineering, Zhejiang University, Hangzhou 310027, China (yuguanding@zju.edu.cn).}
}

\maketitle

\begin{abstract}
It has been a long-held belief that judicious resource allocation is critical to mitigating interference, improving network efficiency, and ultimately optimizing wireless communication performance. The traditional wisdom is to explicitly formulate resource allocation as an optimization problem and then exploit mathematical programming to solve the problem to a certain level of optimality. Nonetheless, as wireless networks become increasingly diverse and complex, e.g., in the high-mobility vehicular networks, the  current design methodologies face significant challenges and thus call for rethinking of the traditional design philosophy. Meanwhile, deep learning, with many success stories in various disciplines, represents a promising alternative due to its remarkable power to leverage data for problem solving. In this paper, we discuss the key motivations and roadblocks of using deep learning for wireless resource allocation with application to vehicular networks. We review major recent studies that mobilize the deep learning philosophy in wireless resource allocation and achieve impressive results. We first discuss deep learning assisted optimization for resource allocation. We then highlight the deep reinforcement learning approach to address resource allocation problems that are difficult to handle in the traditional optimization framework. We also identify some research directions that deserve further investigation.
\end{abstract}


\section{Introduction}
%
%
%
%
\IEEEPARstart{O}{ver} the past few decades, wireless communications have been relentlessly pursuing higher throughput, lower latency, higher reliability, and better coverage. In addition to designing more efficient coding, modulation, channel estimation, equalization, and detection/decoding schemes, optimizing the allocation of limited communication resources is another effective approach \cite{han2008resource}.

From Shannon's information capacity theorem \cite{shannon1948mathematical}, power and bandwidth are the two primitive resources in a communication system. They determine the capacity of a wireless channel, up to which the information can be transmitted with an arbitrarily small error rate. For modern wireless communication systems, the definition of communication resources has been substantially enriched. Beams in a multiple-input multiple-output (MIMO) system, time slots in a time-division multiple access system (TDMA), frequency sub-bands in a frequency-division multiple access system, spreading codes in a code-division multiple access system, and even base stations or backhaul links in virtualized wireless networks all count. Judicious allocation of these communication resources in response to channel conditions and user's quality-of-service (QoS) requirements is critical in wireless system optimization. For example, water-filling power allocation needs to be performed over different subcarriers in an orthogonal frequency division multiplexing (OFDM) system or different channel eigen directions in a MIMO system for capacity maximization. Power and spectrum allocation is important in a cellular network, especially with device-to-device (D2D) underlay, to manage interference and optimize network throughput. In addition to optimizing for traditional physical layer communication metrics, such as capacity maximization \cite{Yu2006dual} and power minimization \cite{Wong1999multiuser}, cross-layer resource allocation takes account of the requirements of upper layers, e.g., delay and fairness, through optimizing properly defined utility functions \cite{Song2005utility,Song2005a,Song2005b}.

Historically, the dominant approach to resource allocation is through mathematical programming, where we optimize one of the design criteria of interest, e.g., sum rate maximization or interference minimization, while imposing constraints on the remaining. Despite the remarkable success in this domain, it turns out that many of the formulated optimization problems are difficult to solve \cite{Luo2008complexity}.
Moreover, with a myriad of new applications to support, conventional methods find it increasingly difficult to balance and model the diverse service requirements in a mathematically exact way. As an example, for ultra-reliable and low-latency communications (URLLC), one of the three major 5G usage scenarios, the definitions of latency and reliability are still subject to debate \cite{Bennis2018ultra}, not to mention a principled approach to provide performance guarantees.
A more flexible framework for wireless resource allocation is thus needed, motivating a departure from the traditional wireless design philosophy.

Recently, adaptation of machine learning tools to address difficult problems in wireless communications and networking has gained momentum, ranging from physical layer design \cite{Oshea2017intro,Dorner2018deep,Ye2018power,ye2019GAN,Samuel2019learning,Farsad2018neural}, resource allocation \cite{Sun2018learning}, networking \cite{Zhang2019DLsurvey}, caching \cite{Chang2018learn}, to edge computing \cite{park2018wireless}.
In fact, the latest cycle of enthusiasm for machine learning is largely triggered by the exceptional performance of deep learning in a broad array of application scenarios.
\textcolor{black}{
Deep learning provides multi-layer computation models that learn efficient representations of data with multiple levels of abstraction from unstructured sources.
It enables a powerful data-driven approach to many problems that are traditionally deemed hard due to, e.g., lack of accurate models or prohibitively high computational complexity.
}
In the wireless resource allocation context, deep learning has been demonstrated to achieve significant performance improvement over conventional methods in several recent works \cite{Lee2018deep, wang2018deep,Nasir2019multi}.
In particular, it has been demonstrated that deep reinforcement learning (RL) is capable of providing a nice treatment of service requirements that are hard to model exactly and thus also not subject to any effective optimization approaches.
\textcolor{black}{
Furthermore, we can exploit the rich expert knowledge in wireless communications developed over the past few decades to complement the data-driven deep learning methods to improve data efficiency via, e.g., deep transfer learning \cite{zappone2019wireless,Zappone2019model}.
}
The goal of this paper is to review some of the most promising results and discuss the principles, benefits, and potential challenges of leveraging deep learning to address wireless resource allocation problems in general with application to vehicular networks as a special example.
Since this is a fast evolving field, we do not attempt to exhaustively cover all research achievements in this area but only highlight those that closely align with our theme.
We refer interested readers to other excellent survey and tutorial papers on various aspects of leveraging learning concepts in the wireless context \cite{Qin2019deep,Gunduz2019machine,Ye2018mlv2x,Liang2018toward,chen2017machine,Chang2018learn,Zhang2019DLsurvey} to get a complete picture.
Compared to them, this paper differentiates itself in that we exclusively focus on wireless resource allocation using deep learning. We emphasize the fundamental properties of this category of problems that make the deep learning approach appealing and demonstrate through extensive examples how to unleash the power of this promising method to its fullest extent.

The paper is organized as follows.
In Section II, we discuss the limitations of traditional optimization methods for wireless resource allocation and motivate deep learning in addressing the issue.
In Section III, we present examples on how to leverage deep learning to solve resource optimization problems in a more efficient way.
In Section IV, deep RL based methods are discussed in detail that warrant a fundamental shift in treating resource allocation problems in a more flexible and effective framework.
In section V, we recognize and highlight several open issues that are worth further investigation. Concluding remarks are finally made in Section VI.

\section{Machine Learning for Resource Allocation}

\begin{figure}[ht]
    \centering
    \includegraphics[width=0.99\linewidth]{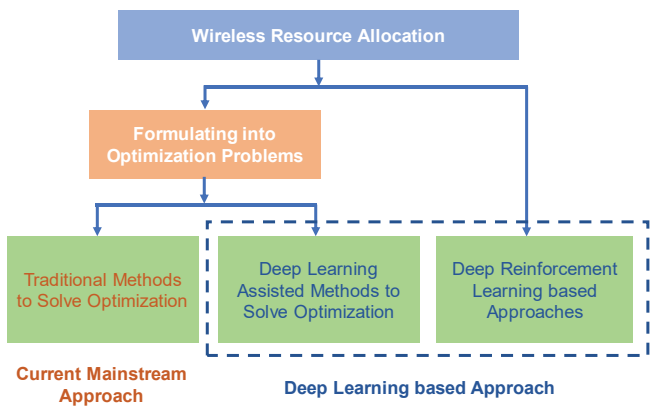}
    \caption{Classification of approaches to wireless resource allocation.}
    \label{fig:sys}
\end{figure}

As shown in Fig.~\ref{fig:sys}, the mainstream approach to wireless resource allocation has long been to formulate the design objective and constraints as an optimization problem. It is then solved to certain levels of optimality, depending on problem complexity and allowable computation time, by leveraging tools from various disciplines, including mathematical programming, graph theory, game theory, etc.
In this section, we discuss the limitations of these optimization approaches and highlight the potentials of uprising data-driven methods enabled by machine learning, in particular deep learning.
Methods that combine the theoretical models derived from domain knowledge and the data-driven capabilities of learning are also briefly discussed.

\subsection{Limitation of Traditional Optimization Approach}
Except in a few simple cases, where we are fortunate enough to end up with convex optimization that admits a systematic procedure to find the global optimum, most optimization problems formulated for wireless resource allocation are strongly non-convex (continuous power control), combinatorial (discrete channel assignment), or mixed integer nonlinear programming (combined power control and spectrum assignment).
For instance, it has been shown in \cite{Luo2008complexity} that the spectrum management problem in a frequency selective channel, where multiple users share the same spectrum, is non-convex and NP-hard. There is no known algorithm that can solve the problem to optimality with polynomial time complexity.
To deal with problems of this kind, we are often satisfied with a locally optimal solution or some good heuristics without any performance guarantee. More often than not, even these suboptimal methods are computationally complex and hard to be executed in real time.
Another limitation of existing optimization approaches points to the requirement of exact models, which tends to abstract away many imperfections in reality for mathematical tractability, and the solution is highly dependent on the accuracy of the models. However, the wireless communication environment is constantly changing by nature and the resultant uncertainty in model parameters, e.g., channel state information (CSI) accuracy, undermines the performance of the optimized solution.

Finally, as wireless networks grow more complex and versatile, a lot of the new service requirements do not directly translate to the performance metrics that the communication community is used to, such as the sum rate or proportional fairness.
For example, in high-mobility vehicular networks, the simultaneous requirements of capacity maximization for vehicle-to-infrastructure (V2I) links and reliability enhancement for vehicle-to-vehicle (V2V) links \cite{Liang2018toward,Ye2018mlv2x} do not admit an obvious formulation. In particular, if we define the reliability of V2V links as the successful delivery of packets of size $B$ within the time constraint $T$ \cite{3GPPr14v2x,Molina2017LTEV}, the problem becomes a sequential decision problem spanning the whole $T$ time steps and is difficult to solve in a mathematically exact manner.
To avoid such difficulties, traditional optimization based methods break down the problem into isolated resource allocation decisions at each time step without considering the long-term effect. For instance, methods in \cite{Liang2017resource,Liang2018graph,Sun2016radio,Sun2016cluster} reformulate the requirement as a signal-to-interference-plus-noise ratio (SINR) constraint at each decision step and then use various optimization techniques to solve for the resource allocation solution. Such a practice loses the flexibility to balance V2I and V2V performance across the whole time $T$ and leads to inevitable performance loss.

\subsection{Deep Learning Assisted Optimization}
Deep learning allows multi-layer computation models that learn data representations with multiple levels of abstraction \cite{lecun2015deep,juang2016deep}.
Each layer computes a linear combination of outputs from the previous layer and then introduces nonlinearity through an activation function to improve its expressive power.
Deep learning has seen a recent surge in a wide variety of research areas due to its exceptional performance in many tasks, such as speech recognition and object detection.
Coupled with the availability of more computing power and advanced training techniques, deep learning enables a powerful data-driven approach to many problems that are deemed difficult traditionally.
In the context of resource allocation, this sheds light on solving hard optimization problems (at least partially).

In the simplistic form, deep learning can be leveraged to learn the correspondence of the parameters and solutions of an optimization problem. The computationally complex procedure to find optimal or suboptimal solutions can be taken offline. With the universal approximation capability of deep neural networks (DNNs), the relation between the parameter input and the optimization solution obtained from any existing algorithm can be approximated. For implementation in real time, the new parameter is input into the trained DNN and a good solution can be given almost instantly, thus improving its potential for adoption in practice.

In learning tasks, the DNN is usually trained to minimize the discrepancy between the output and the ground truth given an input. With this idea in mind, we can leverage deep learning to directly minimize or maximize the optimization objective, i.e., treating the objective as the loss function in supervised learning. Then various training algorithms, such as stochastic gradient descent, can be employed to find the optimization solution.
Compared with the direct input-output relation learning, this approach lifts the performance limit imposed by the traditional optimization algorithms that are used to generate the training data.
Alternatively, deep learning can be embedded as a component to accelerate some steps of a well-behaved optimization algorithm, such as the pruning stage of the branch-and-bound in \cite{Shen2019lorm,Lee2019branch}. This method leverages the theoretical models developed with expert knowledge and achieves near-optimal performance with significantly reduced execution time.

\subsection{Deep Reinforcement Learning based Resource Allocation}

\begin{figure}[ht]
    \centering
    \includegraphics[width=0.8\linewidth]{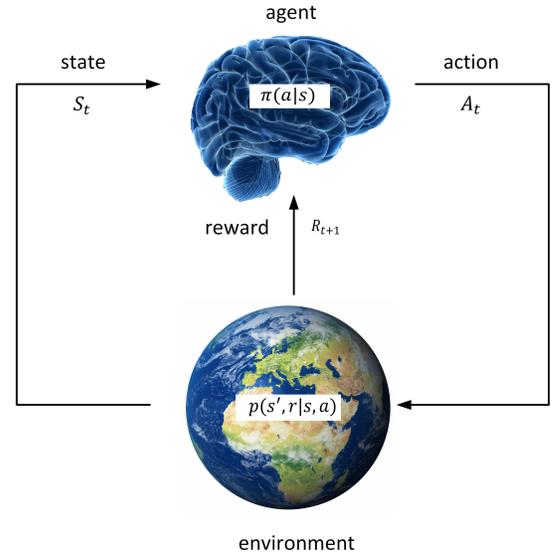}
    \caption{The agent-environment interaction in reinforcement learning. }
    \label{fig:rl}
\end{figure}

RL addresses sequential decision making via maximizing a numeric reward signal while interacting with the unknown environment, as illustrated in Fig.~\ref{fig:rl}.
Mathematically, the RL problem can be modeled as a Markov decision process (MDP). At each discrete time step $t$, the agent observes some representation of the environment state $S_t$ from state space $\mathcal{S}$, and then selects an action $A_t$ from the action set $\mathcal{A}$. Following the action, the agent receives a numerical reward $R_{t+1}$ and the environment transitions to a new state $S_{t+1}$, with transition probability $p(s', r|s,a)$.
In RL, decision making manifests itself in a policy $\pi(a|s) $, which is a mapping from states in $\mathcal{S}$ to probabilities of selecting each action in $\mathcal{A}$.
The goal of learning is to find an optimal policy $\pi_*$ that maximizes the expected accumulative rewards from any initial state $s$.
The RL framework provides native support for addressing sequential decision making under uncertainty that we encounter in, e.g., the resource allocation problem in vehicular networks.
The problem can be tackled by designing a reward signal that correlates with the ultimate objective and the learning algorithm can figure out a decent solution to the problem automatically. Indeed, it is the flexibility of reward design in RL that makes it appealing for solving problems that are deemed difficult traditionally due to the inability to model the design objective exactly.

\textcolor{black}{Here we briefly review two basic RL algorithms, i.e., (deep) Q-learning and REINFORCE, as representatives of the value-based and policy-based RL methods, respectively.
We also provide a concise introduction to the actor-critic algorithm that blends the two RL categories.
As for more advanced algorithms, such as natural policy gradients \cite{kakade2002natural}, trust region policy optimization \cite{schulman2015trust}, async advantage actor-critic \cite{mnih2016asynchronous} and others, we refer interested readers to the given reference papers and the survey on deep RL in \cite{franccois2018introduction,Arulkumaran2017deep}. }

\subsubsection{Q-Learning}
{Q-Learning} is a popular model-free method to solve RL problems \cite{watkins1992q}. It is based on the concept of action-value function, $q_{\pi}(s,a)$, for policy $\pi$, which is defined as the expected accumulative rewards starting from state $s$, taking action $a$, and thereafter following policy $\pi$.
The action-value function of the optimal policy, $q_*(s,a)$, satisfies a recursive relation, known as the Bellman optimality equation.
In principle, one can solve this system of nonlinear equations for $q_*(s,a)$ if the dynamics $p(s',a'|s,a)$ are known. Once $q_*$ is obtained, it is easy to determine the optimal policy $\pi_*(a|s)$ follwing a greedy procedure.
Q-learning avoids the difficulty of acquiring the dynamics by taking an iterative update approach, given by
\begin{align}
Q(S_t, A_t) & \leftarrow  Q(S_t, A_t) \nonumber\\
    + & \alpha\Big[ R_{t+1}  + \gamma \max\limits_{a'} Q(S_{t+1}, a')  - Q(S_t, A_t)\Big],
\end{align}
where $\alpha$ is the step-size parameter, $\gamma\in(0,1]$ is the MDP discounter factor, and the choice of $A_t$ in state $S_t$ follows some soft policies, e.g., the $\epsilon$-greedy, meaning that the action with maximal estimated value is chosen with probability $1-\epsilon$ while a random action is selected with probability $\epsilon$.

\subsubsection{Deep Q-Network with Experience Replay}
In many problems of practical interest, the state and action space can be too large to store all action-value functions in a tabular form. As a result, it is common to use function approximation to estimate these value functions.
In deep Q-learning \cite{mnih2015human}, a DNN  parameterized by $\theta$, called deep Q-network (DQN), is used to represent the action-value function.
The state-action space is explored with some soft policies, e.g., $\epsilon$-greedy, and the transition tuple $(S_{t}, A_t, R_{t+1}, S_{t+1})$ is stored in a replay memory at each time step.
The replay memory accumulates experiences over many episodes of the MDP. At each step, a mini-batch of experience $\mathcal{D}$ are uniformly sampled from the memory for updating $\theta$ with variants of stochastic gradient descent methods, hence the name experience replay, to minimize the sum-squared error:
\begin{align}\label{marl:eq:dqn}
\sum\limits_{\mathcal{D}} \left[R_{t+1} + \gamma\max\limits_{a'}Q(S_{t+1}, a';\theta^-) - Q(S_t, A_t; \theta) \right]^2,
\end{align}
where $\theta^-$ is the parameter set of a target Q-network, which is duplicated from the training Q-network parameters set $\theta$ periodically and fixed for a couple of updates.
Experience replay improves sample efficiency through repeatedly sampling stored experiences and breaks correlation in successive updates, thus also stabilizing learning.

\subsubsection{\textcolor{black}{Policy Gradient and Actor Critic}}
\textcolor{black}{
Different from value-based RL methods that estimate a value function and then use that to compute a deterministic policy, the policy gradient methods directly search the policy space for an optimal one.
The policy is usually represented by a function approximator, such as a DNN, parameterized by $\theta$, i.e., $\pi_\theta(a|s)$.
The parameter set $\theta$ is updated in the direction of improvement of a certain performance measure, $J(\theta)$, by gradient descent,
$\theta \leftarrow \theta + \alpha \nabla J(\theta)$,
with proper step size $\alpha$ \cite{sutton2000policy}. In episodic learning tasks, we define $J(\theta)$ as the expected return from the start state,
\begin{align}\label{eq:J_theta}
    J(\theta) = \mathbb{E}\left\{\sum_{t=0}^T \gamma^t R_{t+1} \right\},
\end{align}
where $T$ is final time step. From the policy gradient theorem \cite{sutton1998reinforcement,sutton2000policy}, the gradient $\nabla J(\theta)$ follows
\begin{align}\label{eq:J_grad}
    \nabla J(\theta) = \mathbb{E}_\pi \left\{  \nabla_\theta \log\pi_\theta(A_t|S_t) q_{\pi_\theta}(S_t,A_t)\right\}.
\end{align}
In the REINFORCE algorithm, we use the accumulative rewards obtained from time step $t$ onward as an unbiased estimate of $q_{\pi_\theta}(S_t,A_t)$ and then run multiple episodes of the task following policy $\pi_\theta(a|s)$ to update the parameter set $\theta$ in each time step $t$ during training.
}
\textcolor{black}{
In fact, the REINFORCE algorithm suffers from high variance and we can simultaneously learn an approximation of the value function for variance reduction, i.e.,
\begin{align}\label{eq:critic}
Q_w(s,a) \approx q_{\pi_\theta}(s,a),
\end{align}
with parameter set $w$ to accelerate learning.
Various policy evaluation methods, such as TD learning \cite{sutton1998reinforcement}, can be leveraged to solve the value approximation problem.
In the actor-critic terminology, the value function approximation $Q_w(s,a)$ is called the critic while the approximate policy $\pi_\theta(a|s)$ is called the actor. Then the actor-critic algorithm follows an approximate policy gradient  determined by \eqref{eq:J_grad} and \eqref{eq:critic}.
Finally, we remark that compared with value-based RL methods, the algorithms involving policy gradients can learn stochastic policies and tend to be more effective in high-dimensional or continuous action space.
Nevertheless, they are more likely to converge to a local rather than global optimum.
}

In fact, there are a good number of existing studies leveraging the concept of MDP for resource allocation, e.g., the delay-optimal OFDMA power control and spectrum allocation in \cite{Cui2010distributive,Lau2010delay} and virtualized radio resource scheduling for software-defined vehicular networks in \cite{Zheng2016delay}. However, we do not treat them as (deep) RL approaches as they assume the availability of MDP transition dynamics more or less and are not learning from interactions with unknown environment. Such trial-and-error learning behavior is a key ingredient in making RL as successful as it is today, from our perspective.

\section{Deep Learning Assisted Optimization for Resource Allocation}
This section deals with the employment of deep learning to find (near-) optimal solutions of the optimization problems for wireless resource allocation in an efficient manner. There are three ways to incorporate deep learning in solving the optimization problems:
\begin{itemize}
    \item The supervised learning paradigm: the DNNs are applied to learn the mapping from the parameter input to the solution of a given optimization algorithm.
    \item The objective-oriented unsupervised learning paradigm: the optimization objective is employed as the loss function, which is optimized directly during training.
    \item {The learning accelerated optimization paradigm: the deep learning technique is embedded as a component to accelerate some steps of a given optimization algorithm. }
\end{itemize}

\subsection{Supervised Learning Approach for Optimization}
The most straightforward way to leverage deep learning for resource allocation is treating a given optimization problem as a black box and using various deep learning techniques to learn its input-output relation.
In this case, a traditional optimization method will act as a supervisor, whose output will serve as the ground truth for training the DNNs.
With the universal approximation ability of DNNs, the mapping from the parameter input to the solution of the given optimization algorithm can be approximated.

\begin{figure}[ht]
\centering
\subfigure[Training stage.]{
			\includegraphics[clip,width=0.8\linewidth]{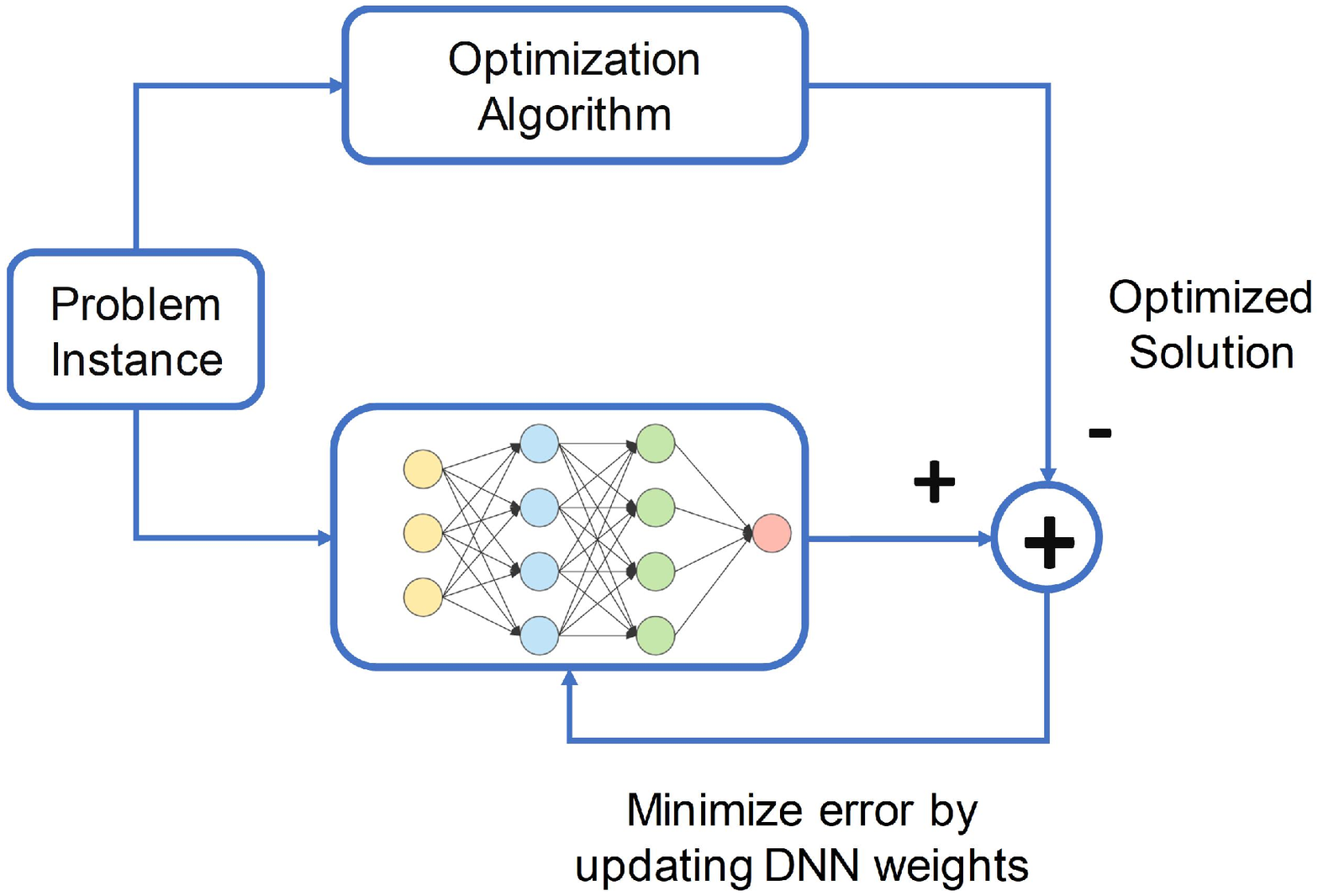}
	}
\subfigure[Testing stage.]{
			\includegraphics[clip,width=0.8\linewidth]{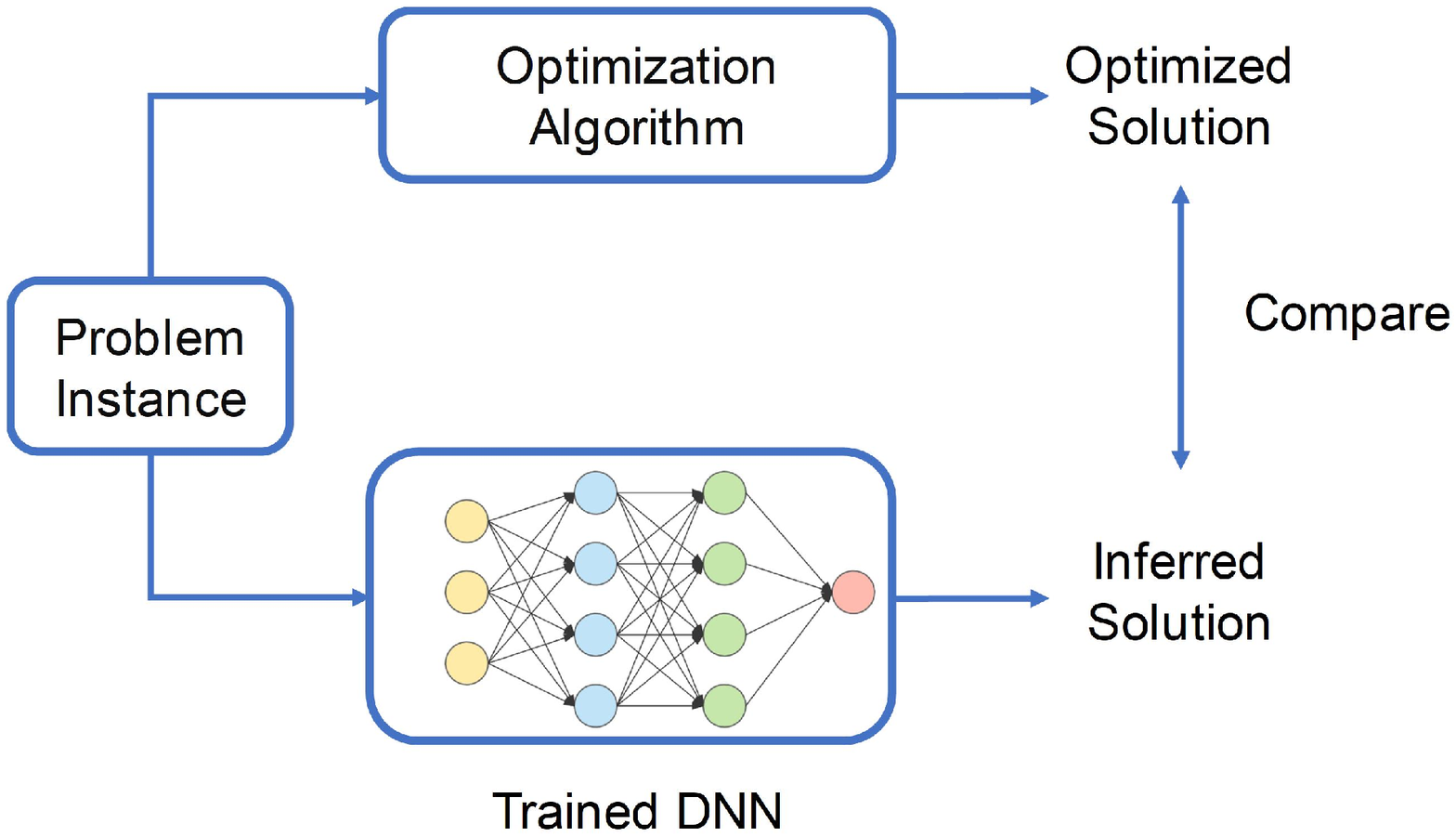}
	}
\caption{The supervised paradigm of using deep learning for optimization.} \label{fig:blackbox}
\end{figure}

The training and testing stages are conveniently shown in Fig.~\ref{fig:blackbox}.
During the training stage in Fig.~\ref{fig:blackbox}(a), labeled samples are generated by running some algorithms of the involved optimization problem using simulated data.
We then leverage the obtained training data set to minimize the discrepancy between DNN outputs and the optimized solutions by updating DNN weights. In the testing stage in Fig.~\ref{fig:blackbox}(b), we also generate the labeled testing samples using the same mathematical algorithm. We pass the parameters of a new problem instance as input to the trained network and collect the inferred solution before comparing it with its corresponding true label.

DNNs have shown their power in mimicking solutions of state-of-the-art optimal or heuristic algorithms while reducing computational complexity.
In \cite{Sun2018learning}, the NP-hard power allocation problem in an interference channel has been solved by a deep learning enabled method that uses a DNN to approximate the weighted minimum mean-squared error (WMMSE) algorithm \cite{Shi2011iteratively}. Simulation results suggest that the performance of DNN is very close to that of the WMMSE algorithm while significantly reducing computational complexity.
\textcolor{black}{
The challenging energy-efficient power control problem in a multi-cell network has been solved by deep learning in \cite{matthiesen2018deep}, which learns from the optimally performing branch-and-bound (B\&B) algorithm. It is worth noting that in \cite{matthiesen2018deep} a specially designed bound has been proposed for the B\&B algorithm to reduce the computational complexity of offline training data generation. The learned DNN is demonstrated to achieve virtually optimal performance in terms of weighted sum energy efficiency while requiring an extremely small online complexity.
}
In \cite{Lee2018lsap}, a similar paradigm is used to deal with a classical combinatorial optimization problem, i.e., the linear sum assignment programming (LSAP) problem, which is frequently encountered in wireless resource allocation. The LSAP problem is about how to assign $n$ jobs to $n$ people in the best way so that some utility (cost) function can be maximized (minimized).
The optimal solution to the LSAP problem can be obtained by the Hungarian algorithm \cite{kuhn1955hungarian} with a computational complexity of $O(n^3)$, which is impractical for real-time implementation in many applications.
To reduce complexity, the LSAP problem has been first decomposed into several sub-assignment problems, which are essentially classification problems. Then, DNNs are utilized to solve each sub-assignment problem by approximating the solutions obtained from running the Hungarian algorithm offline. Finally, a low-complexity greedy collision-avoidance rule has been used to obtain the inferred output for the LSAP problem. As shown in Table~\ref{table1}, DNNs, including the feed-forward network (FNN) and convolutional neural network (CNN), can be used to obtain a real-time solution to the LSAP problem with a slight loss of accuracy.

\begin{table}[ht]
	\small
	\caption{Performance comparison of different methods for LSAP problems with $n=4$.}
	\label{table1}
	\vspace{-1em}
	\centering
	\begin{tabular}{cccc}
		\toprule
		& Hungarian Algorithm & CNN & FNN \\
		\midrule
		Time & $0.5916$ & $0.0120$ & $0.0040$\\
		\midrule
		Accuracy & $100\%$ & $92.76\%$ & $90.80\%$\\
		\bottomrule
	\end{tabular}
\end{table}

\begin{figure}[ht]
	\centering
	\includegraphics[width=0.99\linewidth]{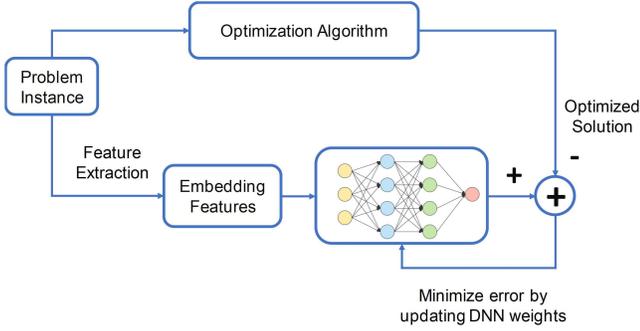}
	\caption{Training stage for the feature embedded paradigm.}
	\label{fig:featureembed}
\end{figure}

In addition, the features of communication networks can be exploited to improve sample efficiency. For instance, a link scheduling method without CSI has been proposed in \cite{cui2018globecom,cui2019spatial,Lee2019graph} for D2D networks with the help of feature embedded paradigm. The link scheduling problem focuses on a densely deployed D2D network with a large number of mutually interfering links. The goal of link scheduling problem is to maximize the network utility by activating a subset of links at any given time.  It can be formulated as a non-convex combinatorial optimization problem. Traditional methods are based on mathematical optimization techniques with the help of accurate CSI. For a network with $N$ D2D links, $N^2$ channel coefficients need to be estimated, which is time- and resource-consuming. In order to bypass the channel estimation stage, the feature embedded paradigm is used for link scheduling. In \cite{cui2018globecom,cui2019spatial}, transmitter and receiver density grids are first constructed for a given D2D network, and then two designed convolutional filters are used to learn the interference pattern among different D2D links. The convolutional stage corresponds to the feature extraction process in Fig.~\ref{fig:featureembed}.
Outputs from the convolutional stage are then input into a DNN that learns from the scheduling results of the state-of-the-art FPLinQ algorithm \cite{Shen2017fplinq} as in \cite{cui2018globecom} or in an unsupervised manner as in \cite{cui2019spatial} to be discussed later. The simulation results suggest that DNNs can effectively learn the network interference topology and perform scheduling to near optimum without the help of accurate CSI. The proposed method needs 800,000 training samples and has good scalability and generalizability to different topologies.

To further reduce the number of required training samples, a graph embedding method has been developed in \cite{Lee2019graph}. For a given D2D network, a graphical model is constructed, where each node corresponds to a D2D link. Then graph embedding is used to learn the vector representation of each node based on topology  information. The output feature vector is then input to a DNN for link scheduling. Compared with the kernel based feature extraction process in \cite{cui2018globecom,cui2019spatial}, the graph embedding process only involves the nonlinear function mapping and can be learned with fewer training samples.
The simulation results are summarized in Table~\ref{table2}. Note that there are two metrics: {classifier accuracy} and {average sum rate}. The former reflects the similarity between the scheduling output of the proposed method and the state-of-the-art FPLinQ algorithm  for link scheduling. The latter is the normalized sum rate achieved by the proposed method with respect to that by the FPLinQ algorithm. From the table, the proposed method in \cite{Lee2019graph} only needs hundreds of training samples to achieve good performance without accurate CSI.

\begin{table}
	\footnotesize
	\caption{Performance of Graph Embedding based Link Scheduling.}
	\label{table2}
	\centering
	\begin{tabular}{|c|c|c|c|c|}
		\hline
		Number of Training Samples  & 200 &500 &1000 &1500\\
		\hline
		Classifier Accuracy& 0.8120 & 0.8208 &0.8192 &0.8388 \\
		\hline
		Average Sum Rate & 0.9362 & 0.9395 &0.9406 &0.9447 \\
		\hline
	\end{tabular}
\end{table}


\subsection{Unsupervised Learning Approach for Optimization} \label{subsec:unsupervised}

The supervised learning paradigm sometimes suffers from several disadvantages.
First, since the ground truth must be provided by conventional optimization algorithms, performance of the deep learning approach will be bounded by that of the conventional algorithms.
Second, a large number of labelled samples are usually required to obtain a good model as demonstrated in \cite{Sun2018learning} and \cite{Lee2018lsap} that use 1,000,000 and 50,000 labelled samples for training, respectively. However, high-quality labelled data are difficult to generate in wireless resource allocation due to, e.g., inherent problem hardness and computational resource constraints. It further limits scalability of the supervised learning methods.


In order to improve performance of the deep learning enabled optimization, unsupervised learning approaches have been proposed to train the DNN according to the optimization objective directly, instead of learning from the conventional optimization approach. In general, deep learning is trained to optimize a loss function via stochastic gradient descent. The loss functions are designed from case to case. For instance, in classification problems, the cross-entropy loss is often used while in regression problems, $l_1$ and $l_2$ losses are preferred.
Therefore, it is natural to use the objective function from the optimization problem as the loss function so that the objective function can be optimized based on the stochastic gradient descent during training.

Various loss functions have been utilized to train the DNNs for wireless resource management and shown to outperform state-of-the-art heuristics.
For example, the sum rate of multiple users is treated as the loss function to train a fully connected DNN in \cite{liang2018towardsOptimal} that learns to solve the non-convex sum rate maximization problem in a multi-user interference channel. With the channel power, $h_{i,j}$, as the input, the DNN training loss is set as the sum rate, given by
\begin{equation}\label{eq:se}
    \eta_{SE} = \sum_{i} \log_2\left(1 + \frac{h_{i, i}P_i}{\sigma^2 + \sum_{k\ne i}h_{k, i}P_k}\right),
\end{equation}
where $P_i$, the $i$th output the DNN, denotes the transmit power of the $i$th transmitter, $\sigma^2$ is the noise power, and $h_{i,j}$ denotes the channel power from transmitter $i$  to receiver $j$.
Furthermore, in \cite{Lee2018deep}, the CNN is used for power control strategy learning in a similar scenario. The DNN can be trained to improve the spectral efficiency as defined in \eqref{eq:se} or the energy efficiency, which is expressed as
\begin{equation}
    \eta_{EE} = \sum_{i} \frac{\eta_{SE}^{(i)}}{P_i + P_c},
\end{equation}
where $\eta_{SE}^{(i)}$ represents the $i$th summation term in \eqref{eq:se} and $P_c$ is the fixed circuit power.
The performance of both unsupervised learning approaches is shown to be better than the state-of-the-art WMMSE heuristic \cite{Shi2011iteratively}.

Along this line of thought, resource allocation is further treated as a functional optimization problem in \cite{Eisen2019learning} that optimizes the mapping from the channel states to power allocation results such that the \emph{long term average} performance of a wireless system is maximized.
To enforce constraints in the learning procedure, such as the power budget and QoS requirements, training of DNNs can be undertaken in the dual domain, where the constraints are linearly combined to create a weighted objective.
Primal-dual gradient descent is then proposed as a model-free learning approach, where the gradients are estimated by sampling the model functions and wireless channels with the policy gradient method.


\textcolor{black}{Conventionally, the loss functions for deep learning training are designed for regression and classification tasks.
These loss functions are usually of simple forms and have theoretical guarantees for convergence.
For instance, the deep linear networks can obtain the global minima with convex loss functions \cite{laurent2018deep} and over-parametrized networks with nonlinear activation functions can achieve zero training loss with the quadratic loss function \cite{du2018gradient}.
In the resource allocation problems, however, the loss functions deviate from the commonly used ones in the regression and classification tasks, and hence issues arise from the theoretical and practical aspects in obtaining good performance.}
As such, in the majority of existing works, additional techniques have been applied to improve training performance for the optimization objectives in wireless resource allocation. For example, in \cite{cui2019spatial}, the DNNs are designed with special structures, including the use of convolutional filters and feedback links.
In \cite{liang2018towardsOptimal}, multiple deep networks are ensembled together for better performance while in \cite{Lee2018deep}, the DNN is pre-trained with the WMMSE solution as the ground truth.

\subsection{Deep Learning Accelerated Optimization}

In the two aforementioned learning paradigms, the optimization procedure is usually viewed as a black box and then completely replaced by a deep learning module. These approaches do not require any prior information about the optimization problems, but need a large amount of training data, labelled or unlabelled, to obtain good performance, albeit with some efforts to improve sample efficiency \cite{cui2018globecom,cui2019spatial,Lee2019graph}.
To address the issue, it has been proposed to leverage the domain knowledge by embedding deep learning as a component to accelerate certain parts of a well-behaved optimization algorithm.

The mixed integer nonlinear programming (MINLP) problem is frequently encountered in wireless resource allocation, which is generally NP-hard and difficult to solve optimally. The generic formulation is given by \cite{Shen2019lorm}
\begin{align}
    & \maximize\limits_{\pmb{\alpha, \omega}} ~ f(\pmb{\alpha}, \pmb{\omega}) \\
    &  \sbto~~ Q(\pmb{\alpha}, \pmb{\omega}) \leq 0, \nonumber\\
    & ~~~~~~~~~~~~~~~  \alpha_i \in \mathbb{N}, \omega_i \in \mathbb{C}, \forall i, \nonumber
\end{align}
where $f(\cdot, \cdot)$ is the objective function, $\alpha_i$ and $\omega_i$ are the elements of $\pmb{\alpha}$ and $\pmb{\omega}$, and $Q(\cdot, \cdot)$ represents certain constraints, such as power or QoS constraints. $\mathbb{N}$ and $\mathbb{C}$ denote some discrete and continuous set constraints, respectively.
Traditional approaches to the MINLP problem are often based on mathematical programming, such as the globally optimal B\&B algorithm, which nontheless has high complexity for real-time implementation.

\begin{table}
	\footnotesize
	\caption{Performance of Accelerated B\&B Algorithm with Different Numbers of Training Samples for the Scenario with 5 cellular users and 2 D2D pairs.}
	\label{table3}
	\centering
	\begin{tabular}{|c|c|c|c|c|}
		\hline
		Number of training samples& 50 & 100 & 150 & 200 \\
		\hline
		Ogap  & 3.88\%& 3.23\%& 2.27\% & 2.01\%\\
		\hline
		Speed& 2.50x& 2.21x& 2.17x& 2.06x\\
		\hline
	\end{tabular}
\end{table}

The resource allocation in cloud radio access networks (RANs) and D2D systems has been studied in \cite{Shen2019lorm} and \cite{Lee2019branch}, respectively, which can be formulated as a MINLP problem and subject to the B\&B algorithm.
By observation, the branching procedure is the most time consuming part in the B\&B algorithm and a good pruning policy can potentially reduce the computational complexity. The more nodes that would not lead to the optimal solution are pruned, the less time is consumed.
Therefore, algorithm acceleration can be formulated into a pruning policy learning task. With invariant problem-independent features and appropriate problem-dependent features selection, the pruning policy learning task can be further converted into a binary classification problem to be solved by deep learning. Simulation results of the learning accelerated optimization method for D2D networks are summarized in Table~III. In the table, ogap, or optimality gap, means the performance gap between the optimal solution and the one achieved by the accelerated algorithm, while speed refers to the speedup with respect to the original B\&B algorithm. From the table, the accelerated B\&B algorithm can achieve near-optimal performance and meanwhile substantially reduce computational complexity using only tens to hundreds of training samples.

\section{Deep Reinforcement Learning Based Resource Allocation}
Deep RL has been found effective in network slicing \cite{li2018deep}, integrated design of caching, computing, and communication for software-defined and virtualized vehicular networks \cite{He2018integrated}, multi-tenant cross-slice resource orchestration in cellular RANs \cite{chen2018multi}, proactive channel selection for LTE in unlicensed spectrum \cite{Challita2018proactive}, and beam selection in millimeter wave MIMO systems in \cite{va2019online}.
In this section, we highlight some exemplary cases where deep RL shows impressive promises in wireless resource allocation, in particular, for dynamic spectrum access, power allocation, and joint spectrum and power allocation in vehicular networks.

\subsection{Dynamic Spectrum Access}

\begin{figure}[ht]
    \centering
    \includegraphics[width=0.9\linewidth]{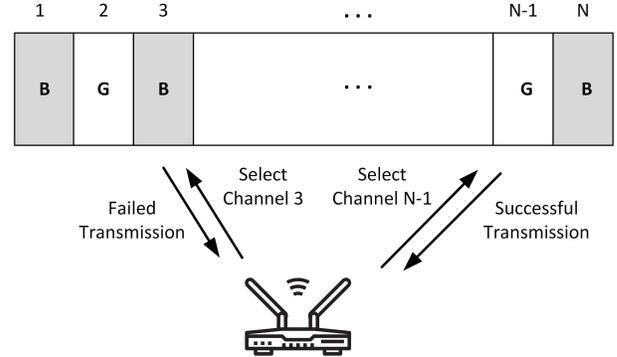}
    \caption{An illustration of dynamic spectrum access of a single transmitter with $N$ channels. ``G'' and ``B'' represent good and bad channel conditions, respectively.}
    \label{fig:dsa}
\end{figure}

In its simplest form, the dynamic spectrum access problem considers a single user that chooses one of $N$ channels for data transmission, as shown in Fig.~\ref{fig:dsa}. In \cite{wang2018deep}, it is assumed that each channel has either ``good'' or ``bad'' conditions in each time slot and the channel conditions vary as time evolves.
The transmission is successful if the chosen channel is good and unsuccessful otherwise. The user keeps transmitting in successive time slots with the objective of achieving as many successful transmissions as possible.

The problem is inherently partially observable since the user can only sense one of the $N$ channels in each time slot while the conditions of other channels remain unknown. To deal with such partial observability and improve transmission success rates, deep RL has been leveraged in \cite{wang2018deep} that takes the history of previous actions and observations (success or failure) up to $M$ time slots as the input to their DQN. The output is a vector of length $N$ with the $i$th element representing the action value of the input state if channel $i$ is selected. The reward design naturally follows: a reward of $+1$ is obtained for successful transmission and $-1$ for failed transmission.
The learning agent employs the $\epsilon$-greedy policy to explore the unknown environment and the accumulated experience is used to improve the policy with the experience replay technique to break data correlation and stabilize training.
When the tested $N=16$ channels are strongly correlated, the proposed learning based dynamic spectrum access method outperforms the implementation friendly model-based Whittle Index heuristic introduced in \cite{Liu2010indexability}. It closely approaches the genie-aided Myopic policy (with channel evolution dynamics known), which is known to be optimal or near-optimal in the considered scenario \cite{zhao2008myopic,Ahmad2009optimality}.

However, as pointed out in \cite{Zhong2018actor}, the DQN enabled approach developed in \cite{wang2018deep} encounters difficulty when the number of channels, $N$, scales large.
To tackle the challenge, a model-free deep RL framework using actor-critic has been further proposed in \cite{Zhong2018actor}. The states, actions, and rewards are similarly defined. But two DNNs are now constructed at the agent, one for the actor that maps the current state (history of actions and observations up to $M$ time slots) to the distribution over all actions and is updated using policy gradient, and the other for the critic that evaluates the value of each action (minus a baseline) under the current policy. The agent explores the unknown environment following the policy defined in the actor network and the accumulated experience is then used to update both the actor (with information from the critic network) and critic networks. Finally, the trained actor network is available to guide the selection of channels at the beginning of each time slot. It has been shown that the actor-critic method has comparable performance as the DQN based approach in \cite{wang2018deep} when $N=16$ and significantly outperforms that when $N$ grows to $32$ or $64$.
Similar to the original DQN based algorithm, the actor-critic approach also demonstrates strong potential to quickly adapt to sudden nonstationary change in the environment while the actor-critic algorithm is more computationally efficient.

The DQN based method is also applied to address dynamic spectrum access in a time-slotted heterogeneous network in \cite{Yu2018deep}, where a wireless device needs to decide whether to transmit in a time slot when it coexists with other devices that follow TDMA and ALOHA protocols. The transmission fails if more than one device attempts to transmit and the goal is also to maximize the number of successful transmissions. Similar to \cite{wang2018deep}, the wireless device constructs a DQN that takes the history of previous actions (TRANSMIT or WAIT) and observations (SUCCESS, COLLISION, or IDLENESS) up to $M$ time slots, as the input, and the output is the value of each action given the current state. The reward is set to $1$ if the observation is SUCCESS and $0$ otherwise.
The learning agent interacts with the unknown wireless environment to gain experience for Q-network training without prior knowledge about the protocols that other devices follow.
With such a model-free deep RL approach that is purely data-driven, the performance is measurably close to the theoretical upper bound that has full model knowledge.

So far, we have demonstrated the power of deep RL in a single user spectrum access problem. In fact, similar observations translate to the multi-user setting as well, albeit with some variations. As investigated in \cite{Naparstek2019deep}, a set of $\mathcal{K} = \{1,\cdots, K\}$  users attempt transmission over $\mathcal{N} = \{1, \cdots, N\}$ orthogonal channels. In each time slot, every user selects a channel to send its data and the transmission is successful (with an ACK signal received) if no others use the same channel.
Different from a single user setting, various design objectives can be defined for multiple users, such as sum rate maximization, sum log-rate maximization (known as proportional fairness), etc., depending on the network utility of interest.
Apart from its strong combinatorial nature, the problem is difficult in that the environment is only partially observable to each user and
nonstationary from user's perspective due to the interaction among multiple users when they are actively exploring and learning.

The deep RL based framework developed in \cite{Naparstek2019deep} assumes a centralized training and distributed implementation architecture, where a central trainer collects the experience from each user, trains a DQN, and sends the trained parameters to all users to update their local Q-networks in the training phase. In the implementation phase, each user inputs local observations into its DQN and then acts according to the network output without any online coordination or information exchange among them. To address partial observability, a long short-term memory (LSTM) layer is added to the DQN that maintains an internal state and accumulates observations over time.
The local observation, $S_i(t)$, of user $i$ in time slot $t$ includes its action (selected channel), the selected channel capacity, and received ACK signal in time slot $t-1$. The observation is then implicitly aggregated in the LSTM layer embedded in the DQN to form a history of the agent.
The action, $a_i(t)$, of user $i$ in time slot $t$ is drawn according to the following distribution to balance exploitation and exploration
\begin{align}
    \text{Pr}\left(a_i(t) = a \right) = \frac{(1-\alpha) e^{\beta Q(a,S_i(t))}}{\sum\limits_{\tilde{a}\in \mathcal{N}} e^{\beta Q(\tilde{a}, S_i(t))} }  + \frac{\alpha}{N+1},
\end{align}
where $\alpha \in (0,1)$, $\beta$ is the temperature to be tuned in training, and $Q(a,S_i(t))$ is the value of selecting channel $a$ for a given observation $S_i(t)$ according to the DQN output.

To address the issue of environment nonstationarity, experience replay, which has been popular in most deep RL training but could continuously confuse the agent with outdated experiences in a nonstationary environment, is disabled during training. More advanced techniques, such as the dueling network architecture \cite{wang2015dueling} and double Q-learning \cite{van2016deep}, are leveraged to improve training convergence.
When compared with the classical slotted ALOHA protocol, opportunistic channel aware algorithm \cite{Cohen2016distributed,To2010exploiting}, and distributed protocol developed in \cite{Hou2014proportionally}, the proposed deep RL method consistently achieves better performance in terms of average channel utilization, average throughput, and proportional fairness with a properly designed reward according to the utility of interest.

\subsection{Power Allocation in Wireless Networks}
Power allocation in wireless networks concerns the adaption of transmit power in response to varying channel and user conditions such that system metrics of interest are optimized.
Consider an interference channel, where $N$ communication links, denoted by $\mathcal{N}=\{1,\cdots,N\}$, share a single spectrum sub-band that is assumed frequency flat for simplicity. The transmitter of link $i$ sends information toward its intended receiver with the power, $P_i$, and the channel gain from the transmitter of link $j$ to the receiver of link  $i$ in time slot $t$ is denoted by  $g_{j,i}^{(t)}$, for any $i, j\in \mathcal{N}$, including both large- and small-scale fading components. Then the received SINR of link $i$ in time slot $t$ is
\begin{align}
    \gamma_i^{(t)}(\mathbf{P}) = \frac{P_i g_{i,i}^{(t)}}{\sum\limits_{j\ne i}P_j g_{j, i}^{(t)} + \sigma^2},
\end{align}
where $\mathbf{P}=\left[P_1,\cdots,P_N \right]^T$ is the transmit power vector for the $N$ links and $\sigma^2$ represents noise power. Then a dynamic power allocation problem to optimize a generic weighted sum rate is formulated as
\begin{align}
    & \maximize\limits_\mathbf{P} ~ \sum\limits_{i=1}^N w_i^{(t)} \cdot \log\left(1+\gamma_i^{(t)}(\mathbf{P}) \right) \label{eq:power} \\
    &  \sbto~~ 0 \le P_i \le P_{\text{max}}, ~ i\in \mathcal{N}, \nonumber
\end{align}
where $P_\text{max}$ is the maximum transmit power for all links and $w_i^{(t)}$ is the nonnegative weight of link $i$ in time slot $t$ that can be adjusted to consider sum rate maximization or proportionally fair scheduling.

Due to the coupling of transmit power across all links, the optimization problem in \eqref{eq:power} is in general non-convex and NP-hard \cite{Luo2008complexity}, not to mention the heavy signaling overhead to acquire global CSI.
To address the challenge, a model-free deep RL based power allocation scheme has been developed in \cite{Nasir2019multi} that can track the channel evolution and execute in a distributed manner with limited information exchange.

The proposed deep RL method in \cite{Nasir2019multi} assumes a centralized training architecture, where each transmitter acts as a learning agent that explores the unknown environment with an $\epsilon$-greedy policy and then sends its exploration experience to a central controller through backhaul links with some delay.
A DQN is trained at the controller using the experience collected from all agents with experience replay, which stores an approximation of action values in different environment states.
After training for a while, the updated DQN parameters are broadcast to all agents that use the parameters to construct/update their own DQNs for distributed execution.
To have a better representation of the communication environment, the state observed by each agent is constructed to include useful local information (such as its own transmit power in the previous time slot, total interference power, and its own channel quality), interference from close interfering neighbors, and its generated interference toward impacted neighbors.
With a proper reward design that targets system-wide performance, the proposed deep RL method learns to adapt the transmit power of each link only using the experience obtained from interaction with the environment. It has been shown to outperform state-of-the-art, including the WMMSE algorithm in \cite{Shi2011iteratively} and fractional programming (FP) algorithm in \cite{Shen2018fractional}, which are assumed to have accurate global CSI.
Remarkably, the developed power allocation method leveraging deep RL returns solutions better and faster without assuming prior knowledge about the channels, and thus can handle more complicated but practical nonidealities of real systems.
Another interesting observation is that the proposed learning based approach shows impressive robustness in the sense that DQNs trained in a different setting (different initialization or numbers of links) can still achieve decent performance. It suggests using a DQN trained on a simulator to jump-start a newly added node in real communication networks is very likely to work in practice.
An extension has been developed in \cite{meng2018power} that considers the case where each transmitter serves more than one receiver and the data-driven deep RL based approach has been demonstrated to achieve better performance than benchmark algorithms.

The power allocation problem has also been studied in \cite{Ghadimi2017reinforcement,Calabrese2018learning}, which consider a cellular network with several cells and the base station in each cell serves multiple users. All base stations share the same frequency spectrum, which is further divided into orthogonal sub-bands within the cell for each user, i.e., there exists inter-cell interference but no intra-cell interference.
Each base station is controlled by a deep RL agent that takes actions (i.e., performs power adaptation) based on its local observation of the environment, including cell power, average reference signal received power, average interference, and a local cell reward. The power control action is discretized to incremental changes of $\{0,\pm 1,\pm 3\}$ dB and the reward is designed to reflect system-wide utility to avoid selfish decisions. The agents take turns to explore the environment to minimize the impact on each other, which stabilizes the learning process. Each deep RL agent trains a local Q-network using the experience accumulated from its interaction with the communication environment.
The deep RL based approach learns to control transmit power for each base station that achieves significant energy savings and fairness among users in the system.

\textcolor{black}{
Power allocation in a more complicated multi-cell network has been considered in \cite{meng2019power}, where each cell has one base station serving multiple users and all base stations transmit over the same spectrum, generating both intra- and inter-cell interference.
Each base station is controlled by an RL agent that decides its transmit power to maximize the network sum throughput. Similar to \cite{Nasir2019multi}, a single RL agent is trained in a centralized manner using experience collected from all transmitters. The trained policy is then shared among all agents and executed in a distributed way.
Deep Q-learning, REINFORCE, actor-critic deep deterministic policy gradient (DDPG) algorithms have been leveraged to train the RL agent and all of them outperform the benchmark WMMSE \cite{Shi2011iteratively} and  FP algorithms \cite{Shen2018fractional}. Among the three algorithms, the actor-critic DDPG is the best in terms of both sum rate performance and robustness.
}

\subsection{Joint Spectrum and Power Allocation: Application Example in Vehicular Networks}

\begin{figure}[ht]
    \centering
    \includegraphics[width=0.9\linewidth]{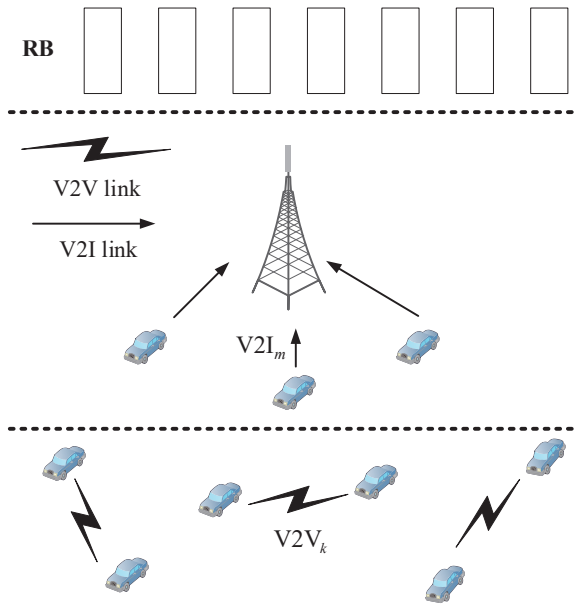}
    \caption{An illustration of spectrum sharing in vehicular networks, where V2V and V2I links are indexed by $k$ and $m$, respectively, and each V2I link is preassigned an orthogonal RB. }
    \label{fig:v2x}
\end{figure}

In a vehicular network as illustrated in Fig.~\ref{fig:v2x}, $K$ V2V links  share the spectrum of $M$ V2I links to improve its utilization efficiency. The V2I links are designed for high data rate entertainment services while the V2V links need to support reliable dissemination of safety-critical messages, formally stated as the successful delivery of packets of size $B$ within the time constraint $T$.
Such a reliability requirement of V2V links is hard to handle with traditional optimization approaches due to its exponential complexity in the length of $T$.  However, it has been shown in \cite{Ye2019deep,Liang2019jsac} that we can nicely treat the issue in the deep RL framework through properly designing a reward that correlates with the objective.
We assume that each V2I link has been assigned an orthogonal resource block (RB) and uses a fixed transmit power.
Then, each V2V transmitter needs to carefully select the V2I RB to share and adjust its transmit power to avoid strong interference and ensure both V2I and V2V links achieve their respective goals.

\subsubsection{Single-Agent RL}
In view of the difficulty to collect global CSI at a central controller in real time, a distributed resource allocation algorithm has been developed in \cite{Ye2019deep} that leverages deep RL.
In particular, each V2V transmitter serves as a learning agent that occupies a local copy of a DQN and follows the $\epsilon$-greedy policy to explore the unknown environment. The observation of each V2V agent represents its own perception of the unknown environment state, given by
\begin{align}\label{eq:stateSARL}
    S_t = \{\mathbf{G}_t, \mathbf{H}_t, \mathbf{I}_{t-1}, \mathbf{N}_{t-1}, L_t, U_t \},
\end{align}
where $\mathbf{G}_t$ and $\mathbf{H}_t$ represent the current V2V signal channel strength and the interference channel strength from the V2V transmitter to the base station over all RBs, respectively, $\mathbf{I}_{t-1}$ is the received interference power, $\mathbf{N}_{t-1}$ is the selected RBs of neighbors in the previous time slot over all RBs, $L_t$ and $U_t$ denote the remaining load and time to meet latency constraint from the current time slot, respectively.
The action of each V2V agent amounts to a selection of RB as well as discrete transmit power levels.
The reward balances V2I and V2V requirements, given by
\begin{align}\label{eq:reward_sarl}
    r_t = \lambda_c \sum\limits_m C^c[m] + \lambda_v \sum\limits_k C^v[k] -\lambda_p(T - U_t),
\end{align}
where $C^c[m]$ and $C^v[k]$ represent the capacity of V2I link $m$ and V2V link $k$, respectively. $\lambda_c, \lambda_v$ and $\lambda_p$ are nonnegative weights to balance different design objectives. In particular, the inclusion of $T - U_t$ in the reward constantly reminds agents of the upcoming deadline for V2V payload transmission and effectively helps improve payload delivery rates for V2V links.

\begin{figure}[ht]
    \centering
    \includegraphics[width=0.99\linewidth]{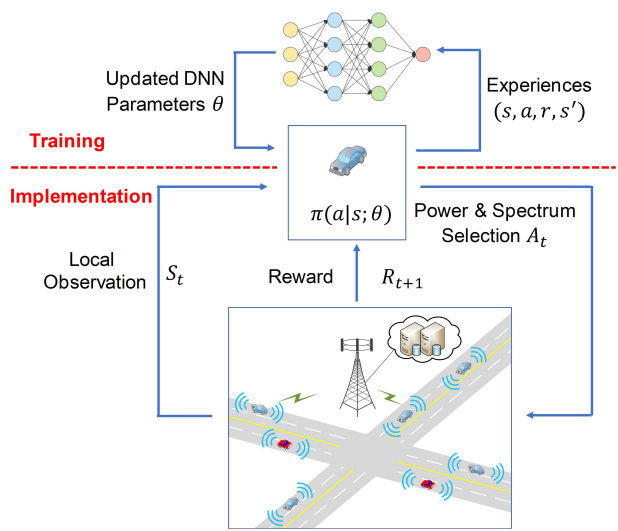}
    \caption{The deep RL training and implementation architecture for resource allocation in vehicular networks. }
    \label{fig:trainArct}
\end{figure}

The reward design in \eqref{eq:reward_sarl} facilitates system-wide performance improvement. But it also dictates the use of a centralized training architecture, where a central controller collects experiences from all V2V agents and compiles the reward for the DQN training.
The system architecture is illustrated in Fig.~\ref{fig:trainArct}.
In the implementation, each V2V agent performs a local observation of the environment and then uses its local copy of the trained DQN to guide its RB selection and power control in a distributed manner.
To alleviate the impact of environment nonstationarity due to mutual interaction among multiple V2V links, the V2V agent takes turns to change its action to stabilize training, rather than acting simultaneously. Experimental results demonstrate that the proposed deep RL based approach can learn from scratch to perform intelligent spectrum and power allocation that outperforms the benchmarks, including the distributed algorithm in \cite{ashraf2016dynamic} and a random baseline, in terms of both sum V2I rate as well as the delivery rate of V2V payloads.

\subsubsection{Multi-Agent RL}
To further improve the network performance and better handle the dynamics in a vehicular network, we investigate in \cite{Liang2019jsac} a multi-agent RL based approach to enable all V2V agents to perform resource allocation simultaneously.
We have iterated throughout the article that such simultaneous actions of all learning agents tend to make the environment observed by each agent highly nonstationary and compromises stability of DQN training.
To address the issue, either the experience replay technique that is central to the success of deep RL is disabled as in \cite{Naparstek2019deep}, or the agents take turns to update their actions as in \cite{Nasir2019multi,Ghadimi2017reinforcement,Ye2019deep}.

We believe that such a turn-taking action update constraint leads to inevitable suboptimality as it is a subset of the simultaneous action space and the disabling of highly efficient experience replay techniques is undesirable.
In response, we leverage the fingerprint based method proposed in \cite{Foerster2017stabilising} that identifies and addresses the source of nonstationarity: the policy change of other agents due to learning. As such, the environment observed by each agent can be made stationary by conditioning on other agents' policy change, i.e., we augment the observation of each V2V agent with an estimate of the policy change of all other agents, the idea of hyper Q-learning \cite{tesauro2004extending}. Further analysis reveals that the agents' policy varies along the learning process, whose trajectory can be tracked by a low-dimensional fingerprint, including the training iteration number $e$ and the probability of selecting a random action, $\epsilon$, in the $\epsilon$-greedy policy. Then we revise the observation of each V2V agent as
\begin{align}
    Z_t^{(k)} = \{S_t^{(k)}, \epsilon, e\},
\end{align}
where $S_t^{(k)}$ contains similar local observations (measurements) of the $k$th V2V agent as in \eqref{eq:stateSARL}.
In addition, we revise the reward design to approach the V2V payload delivery reliability more closely by setting the V2V related reward component at each time step to the sum V2V rate when payload delivery is not finished and to a constant number $\beta$ larger than the largest sum V2V rate when finished.
The design of $\beta$ reflects the tradeoff between designing purely toward the ultimate goal and learning efficiency. For pure goal-directed consideration, we would set the V2V related reward to $0$ at each step until the payload is delivered when the reward is $1$. However, receiving such delayed rewards decreases training efficiency and hence we impart our domain knowledge by incorporating the sum V2V rate as aggregated rewards.
Again we employ the centralized training and distributed implementation architecture in Fig.~\ref{fig:trainArct} to train multiple DQNs for V2V agents with experience replay and then deploy them after training is done.

\begin{figure}[!t]
\centering
\subfigure[V2V transmission rates of MARL.]{\includegraphics[clip,width=0.8\linewidth]{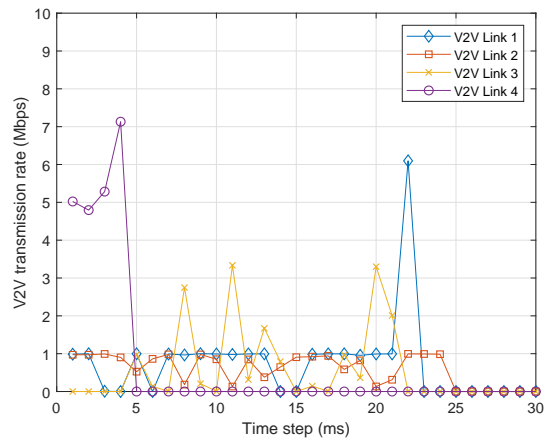}}
\subfigure[V2V transmission rates of the random baseline.]{\includegraphics[clip,width=0.8\linewidth]{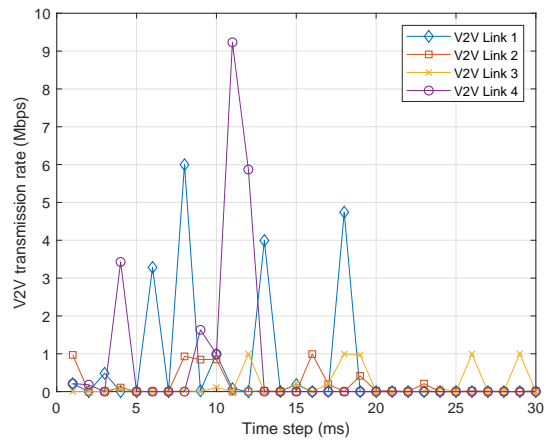}}
\caption{V2V transmission rates of different resource allocation schemes within one episode. Only the results of the beginning $30$ ms are plotted for better presentation. The initial payload size $B = 2,120$ bytes.}\label{fig:v2vRate}
\end{figure}

Remarkably, such a simple approach proves very effective in stabilizing DQN training and combined with the revised reward design, it significantly outperforms both the single-agent RL based approach in \cite{Ye2019deep} and a random baseline. In fact, the proposed multi-agent RL based method encourages V2V cooperation despite the distributed implementation without any online coordination.
To illustrate, we break down in Fig.~\ref{fig:v2vRate} the change of V2V transmission rates of the proposed method and random baseline over the time constraint $T=100$ ms for one episode where all V2V links successfully deliver payload using the proposed method. But for the random baseline, Link 3 fails the task and all others succeed.
Comparing Figs.~\ref{fig:v2vRate}(a) and (b), one can tell that the proposed method enables Link 4 to transmit at a high rate to finish delivery early and Link 1 is restricted to a low rate at the beginning to protect the more vulnerable Links 2 and 3 that achieve decent rates to deliver their payload. Upon successful delivery of Links 2 and 3, Link 1 leverages its good channel quality to quickly deliver its data. Moreover, Links 2 and 3 figure out a clever strategy of alternating transmission to avoid strong mutual interference. In contrast, the random baseline does not enjoy such intelligence and the more vulnerable Link 3 fails the task finally.

We carry the multi-agent RL idea further in \cite{wang2019NIPS}, where we restructure our network architecture substantially to enable a centralized decision making yet with very low signaling overhead. In particular, each V2V transmitter constructs a DNN that learns to compress its local observation (measurement), which is then fed back to the central base station to serve as the input of the stored DQN. The output of the DQN at the base station is the value of each action vector that contains the actions of all V2V transmitters given the current compressed observation.
Then the resource allocation decision made at the central base station is broadcast to all V2V transmitters during implementation, incurring small overhead.
The proposed architecture aligns with the single meta-agent idea of dealing with dynamics in multi-agent RL and has been shown to achieve about $95\%$ of the optimal performance that is obtained by time-consuming brute force search when each local observation is compressed to $36$ bits per time step.


\section{Open Issues and Future Research}\label{sec:open}
Despite many benefits of using deep learning for wireless resource allocation, blindly applying the technique in the wireless environment is insufficient and more deliberation is needed.
In this section, we identify major roadblocks and highlight some research opportunities.

\subsection{Tailoring Deep Learning Architecture for Wireless}
In deep learning, it is well established that different neural network architectures have varied strengths and weaknesses in different application domains. No single architecture can be the best player in all tasks. For instance, CNNs and deep residual networks (ResNets) have excellent performance in image recognition while recurrent neural networks (RNNs) prevail in sequential data processing, such as speech and language.
Likewise, wireless resource allocation has its own unique characteristics that are worth considering when we adapt or redesign an appropriate DNN.
For example, in power allocation, the adjustment of one transmitter's power affects not only its intended receiver but other closely located wireless links, making the problem non-convex. However, if we convert it to the dual domain, solving the problem indeed becomes much easier and the duality gap is shown to be small in most practical cases \cite{Yu2006dual}.
Such domain knowledge is expected to be very helpful for guiding DNN design in the wireless context. But it is not clear what is the best way to use it. In the unsupervised approaches of using deep learning for optimization mentioned earlier, we treat the resource allocation objective as the loss function for learning. More often than not, the objective is different from what we normally use in DNN training, like mean-squared error or cross-entropy. It is unclear whether the existing network architecture best suits the needs of resource allocation objective minimization or maximization. Furthermore, theoretical understanding of the convergence properties of training DNNs with the new objectives and efficient training techniques largely remain unknown.




\subsection{Bridging the Gap between Training and Implementation}
Most, if not all, proposed learning algorithms in the existing literature are trained and tested on an offline simulator. We understand this makes algorithm design and testing quick and easy, and the policy learned from simulated experiences can be used as a jump-starter for real-time deployment, following the idea of transfer learning \cite{zappone2019wireless}.
Nonetheless, it is extremely difficult to build a simulator with high enough fidelity that guarantees the learned policy work can as expected when interacting with the real-world environment.
It seems that the only way we can provide an ultimate answer to the puzzle is to perform policy learning in the real-world. The issue is then how to avoid catastrophic actions while the agents are actively exploring the environment and no concrete policies have been obtained yet.
A seemingly good answer might be to use expert human knowledge to help confine the exploration space \cite{Calabrese2018learning} and guide the learning agent's search within the space. But exactly how to implement the concept in algorithm design with performance guarantee is unclear and worth further investigation.

\subsection{Multi-Agent Consideration in Deep RL}
In the wireless domain, most scenarios that we are interested in are of multi-user nature, whether it is the power control for multiple users within a cell, or joint spectrum and power allocation for multiple V2V links in a vehicular network.
In these cases, actions of one user impact the performance of not only itself but also others nearby. From the perspective of each user, the environment that it observes then exhibits nonstationarity when other users are actively exploring the state and action space for policy learning.
Meanwhile, each user can only obtain local observations or measurements of the true underlying environment state, which, in the deep RL terminology, is a partially observable MDP. Then the learning agent needs to construct its belief state from the previous actions and local observations to estimate the true environment state, which is a challenging issue.
A partial solution to the problem might be to enable inter-agent communications to encourage multi-user coordination and increase local awareness of the global environment state.
That said, such a combination of environment nonstationarity and partial observability makes learning extremely difficult, which is made even worse if we scale the number of users large as in the upcoming internet of things era.
\textcolor{black}{
While there have been some recent works \cite{sharma2019distributed} that attempt to solve the issue and establish the convergence guarantee of mean-field multi-agent RL, more investigation in this direction is still desired.
}

\section{Conclusion}
In this article, we have provided an overview of applying the burgeoning deep learning technology to wireless resource allocation with application to vehicular networks.
We have discussed the limitations of traditional optimization based approaches and the potentials of deep learning paradigms in wireless resource allocation.
In particular, we have described in detail how to leverage deep learning to solve difficult optimization problems for resource allocation and deep RL for a direct answer to many resource allocation problems that cannot be handled or even modeled in the traditional optimization framework.
We have further identified some open issues and research directions that warrant future investigation.

\end{document}